\documentclass[a4paper]{spie}  

\usepackage{amsmath,amsfonts,amssymb}
\usepackage{graphicx}
\usepackage{dcolumn}
\usepackage{bm}
\usepackage{multirow}
\usepackage{tikz}
\usepackage{comment}
\usepackage{siunitx}
\usepackage[colorlinks=true, allcolors=blue]{hyperref}
\PassOptionsToPackage{hyphens}{url}
\newcommand{\subfigimg}[4][,]{%
  \setbox1=\hbox{\includegraphics[#1]{#3}}
  \leavevmode\rlap{\usebox1}
  \rlap{\hspace*{#4}\raisebox{\dimexpr\ht1-2\baselineskip/2}{#2}}
  \phantom{\usebox1}
}

\title{Sensitivity of Transition-Edge Sensors to Strong DC Electric Fields}

\author[a]{K. M. Patel}
\author[a]{D. J. Goldie}%
\author[b]{S. Withington}%
\author[a]{C. N. Thomas}%
\affil[a]{Cavendish Laboratory, University of Cambridge, 
JJ Thomson Avenue, Cambridge CB3 0HE, United Kingdom}%
\affil[b]{Department of Physics, University of Oxford, Oxford OX1 3PU, United Kingdom}
\authorinfo{Further author information: (Send correspondence to K. M. P)\\K. M. P.: E-mail: kmp57@cam.ac.uk}

\begin{document} 
\maketitle

\begin{abstract}
Transition-edge sensors (TESs) have found a wide range of applications in both space- and land-based astronomical photon measurement and are being used in the search for dark matter and neutrino mass measurements. A fundamental aspect of TES physics that has not been investigated is the sensitivity of TESs to strong DC electric fields (10\,\si{\kilo\volt\per\meter} and above).  
Understanding the resilience of TESs to DC electric fields is essential when considering their use as charged particle spectrometers, a field in which TESs could have an enormous impact. Techniques such as x-ray photoelectron spectroscopy produce a high number of low-energy electrons that are not of interest and can be screened from the detector using electrostatic deflection. The use of strong electric fields could also provide a mass-efficient route to prevent secondary electron measurements arising from cosmic radiation in space-based TES applications. Integrating electron optics into the TES membrane provides an elegant and compact means to control the interaction between charged particles and the sensor, whether by screening unwanted particles or enhancing the particle absorption efficiency but implementing such techniques requires understanding the sensitivity of the TES to the resulting electric fields.
In this work, we applied a uniform DC electric field across a Mo/Au TES using a parallel pair of flat electrodes positioned above and below the TES. The electric field in the vicinity of the TES was enhanced by the presence of silicon backing plate directly beneath the TES. Using this arrangement, we were able to apply of electric fields up to 90\,\si{\kilo\volt\per\meter} across the TES. We observed no electric field sensitivity at any field strength demonstrating the capability to use TESs in environments of strong electric fields.
\end{abstract}


\section{Introduction}\label{sec:Intro}
Transition-edge sensors (TESs) are highly-sensitive superconducting energy detectors with applications spanning astronomical instrumentation \cite{gottardi_development_2016,lee_soft_2019,nagler_transition-edge_2021}, single photon detection\cite{hattori_optical_2022,helversen_quantum_2019}, dark matter searches\cite{rothe_tes-based_2018,edelweiss_collaboration_search_2022} and neutrino mass measurements\cite{puiu_transition-edge_2020}. There is also growing interest in using TESs as massive particle calorimeters for electron or ion measurements \cite{rajteri_tes_2020,smith_microcalorimetry_2020,patel_simulation_2021,croce_superconducting_2011}. With the expanding range of applications for this technology, it is important to be aware of how the environments these detectors are used in may impact their performance. The sensitivity of TESs to magnetic fields has been investigated \cite{claycomb_superconducting_1999,harwin_microscopic_2021} but there has been little consideration of the effect of static electric fields on TES operation. Understanding the sensitivity of TESs to DC electric fields is valuable when considering the application of these devices to measuring charged particles, where the ability to use strong electric fields to accelerate, screen or deflect charged particles would be highly advantageous.

Transition-edge sensors exploit the extremely sharp resistance-temperature dependence of a metal in its superconducting transition window to provide high-resolution energy measurement of incident particles. In operation, the TES is voltage-biased at a point within the resistive-superconducting transition. An electrothermal feedback process maintains the device temperature by balancing absorbed power against reduced Ohmic power dissipation. The corresponding change in the device current can be measured to determine the total energy absorbed. A key factor in the versatility of TESs is the ability to decouple energy absorption and measurement into separate absorber and superconducting structures. This allows for a range of absorber designs, targeting specific particles and energy ranges, all using the same detector technology. 

However, as the TES is sensitive to the deposition of any source of thermal energy into the detector, there remains the potential for the absorption and measurement of unwanted particles. For example, in space-based measurements, cosmic ray events have been observed to be a considerable source of measurement noise \cite{planck_hfi_core_team_planck_2011}. Alternatively, in electron spectroscopy techniques such as Auger and x-ray photoelectron spectroscopy, a high level of low-energy electrons under 200\,\si{\electronvolt} are present that are of no interest to the measurement. 

One possible solution to preventing charged particles reaching the detector would be to integrate electrostatic deflection into the TES absorber design itself but this would require the TES to be robust in the presence of the resulting electric field. Such designs could also prove valuable for preventing the measurement of secondary electrons from cosmic rays, for example. Alternatively, instead of screening charged particles, static electric fields could be used to controllably shift the energy of charged particles reaching the TES, allowing one to shift the energy range of interest for a particular experiment to align with the energy range of greatest detector energy resolution. 

It is not immediately clear whether static electric fields with the intensity required for these applications would observably change the behaviour of TESs. While we expect that applying electric fields will have little effect upon the superconducting components themselves, the electric fields could alter the operation of a TES measurement, for example by coupling to two-level systems within the substrate changing the TES heat capacity. There also remains the possibility of coupling between the TES and the field by previously unconsidered mechanisms. Therefore, we investigated the electric field sensitivity of Mo/Au bilayer TESs to determine whether the application of static electric fields to these devices observably alter the behaviour of these devices.

\section{Experimental}\label{sec:Exp}
\begin{figure}[ht!]
\centering
\begin{tabular}[c]{@{}p{0.4\linewidth}@{}p{0.5\linewidth}}
    \subfigimg[height=5.5cm]{a)}{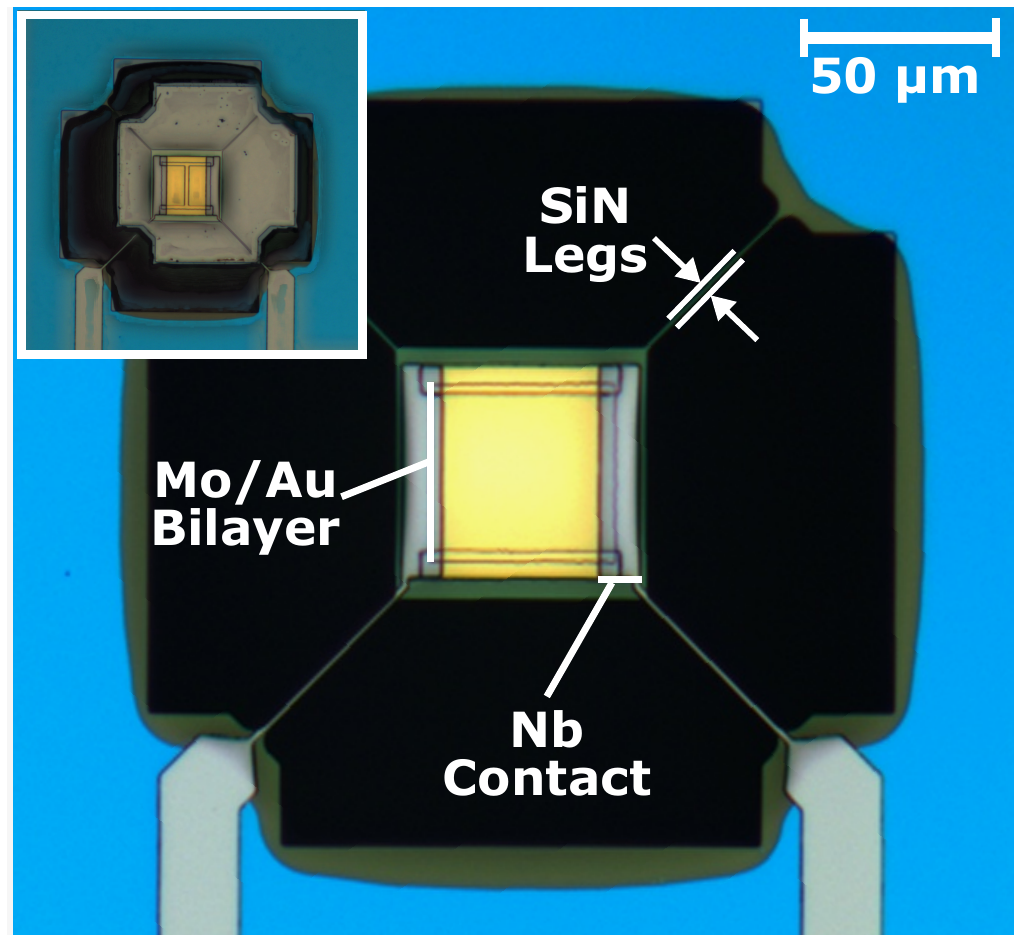}{-12pt} &
    \subfigimg[height=5.5cm]{b)}{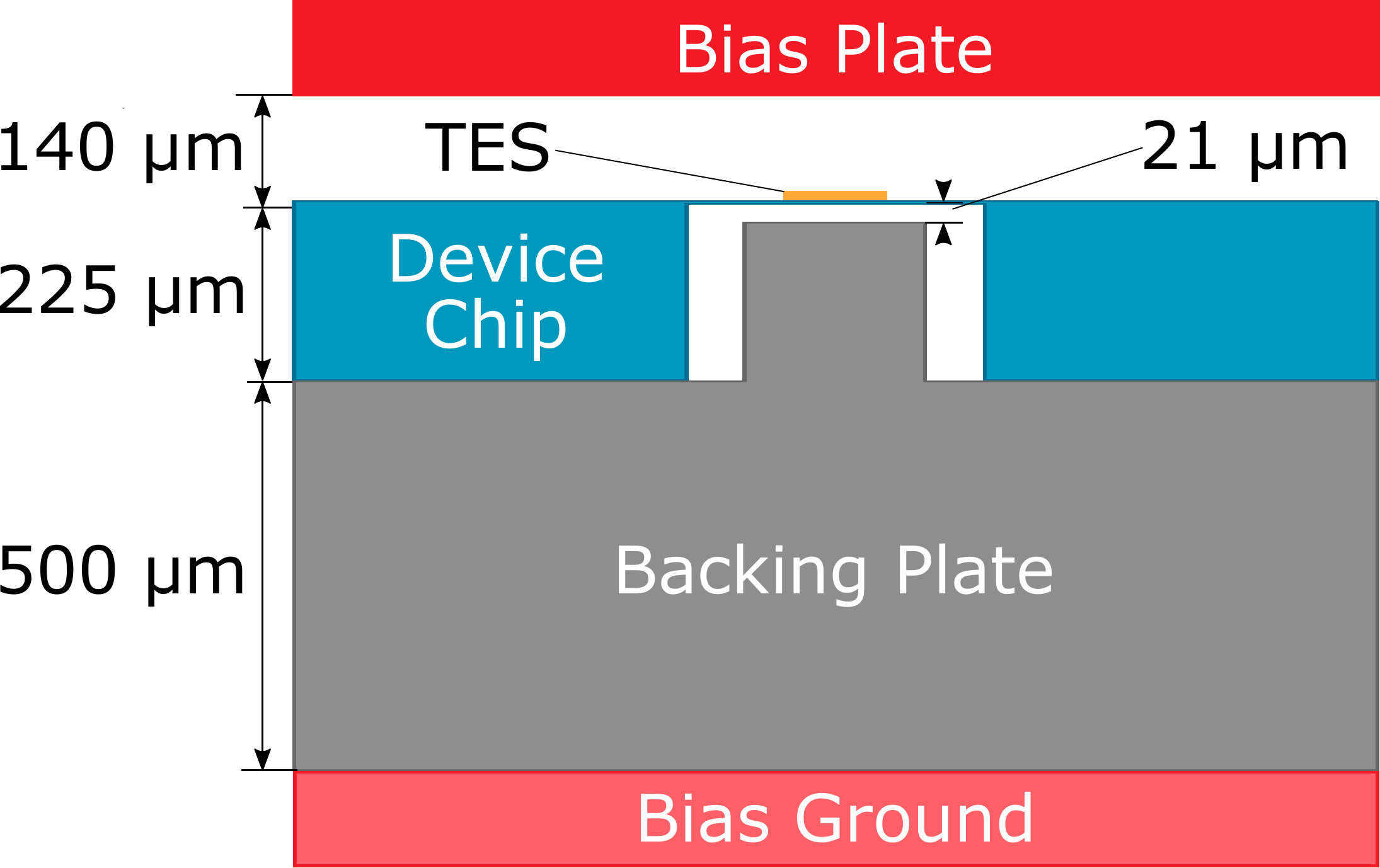}{-12pt}
\end{tabular}

\caption{a) Image of a suspended Mo/Au TES with backing plate pillar visible beneath, representative of design of the 10\(\times\)10\,\si{\micro\meter} device reported here. The inset shows an enhanced depth of field image of a different device with a backing plate pillar visible underneath. b) Diagram of device and backing plate positions in relation to bias plate and ground.}
\label{fig:apparatus}
\end{figure}

TES electric field sensitivity was measured using voltage-biased 40/120\,\si{\nano\meter} Mo/Au bilayer TESs and a \(\text{T}_\text{c}\) of 200\,\si{\milli\kelvin}. The TES was suspended on 225\,\si{\nano\meter} thick micromachined silicon nitride membrane connected to the thermal bath by four 50\,\si{\micro\meter} long and 1.5\,\si{\micro\meter} wide silicon nitride legs. Electrical contacts were made to the TES using niobium leads. Fig.\,\ref{fig:apparatus}a provides an image of a representative TES; the device reported here has bilayer dimensions of 10\(\times\)10\,\si{\micro\meter}. Further details of device design and fabrication are provided in Ref.\,\citeonline{harwin_microscopic_2021}.

The devices were cooled using an adiabatic demagnetisation refrigerator (ADR) with the sample temperature regulated to 120\,mK for the described measurements. TES readout was performed using two-stage SQUID amplification.

The TES device chip was clamped to a grounded copper surface of the ADR cold stage with a backing plate placed between the chip and the grounded circuit. A copper bias plate was attached to the clamp such that it was held 140\,\si{\micro\meter} above device chip, parallel to the TESs. The backing plate was made of silicon with pillars extending 204\,\si{\micro\meter} into each device well, positioning the top of the pillar 21\,\si{\micro\meter} from the device layer. This arrangement is shown in Fig.\,\ref{fig:apparatus}b and an example of a backing plate pillar is visible in Fig.\,\ref{fig:apparatus}a. The top of each silicon pillar was capped with 300\,\si{\nano\meter} of niobium. This experimental arrangement was adapted from \cite{harwin_microscopic_2021} to perform these electric field measurements.

The dielectric backing plate pillars have the effect of enhancing the electric field strength in the vicinity of the device. Modelling the structure with parallel-plate capacitance between the bias and backing plates, the electric field strength experienced by the TES, \(E\), is 
\begin{equation} \label{eq1}
E = V\frac{\epsilon_2}{\epsilon_1d_2+\epsilon_2d_1},
\end{equation}
where \(V\) is the applied plate bias, \(\epsilon_1\) and \(\epsilon_2\) are the permittivities of vacuum and silicon, and \(d_1 = 704\,\mu\)m and \(d_2 = 161\,\mu\)m are the distances from the top of the backing plate pillar to the bias plate and bias ground, respectively. Using Eq.\,\ref{eq1}, at 20\,\si{\volt} applied plate bias and relative permitivitty of 11.5 for low-temperature silicon, the TES experiences 90\,\si{\kilo\volt\per\meter} electric field strength.

\section{Results and Discussion}\label{sec:Results}
\begin{figure}[ht]
\centering
  \begin{tabular}[c]{@{}p{0.45\linewidth} @{}p{0.05\linewidth} @{}p{0.45\linewidth}}
    \subfigimg[height=5.5cm]{a)}{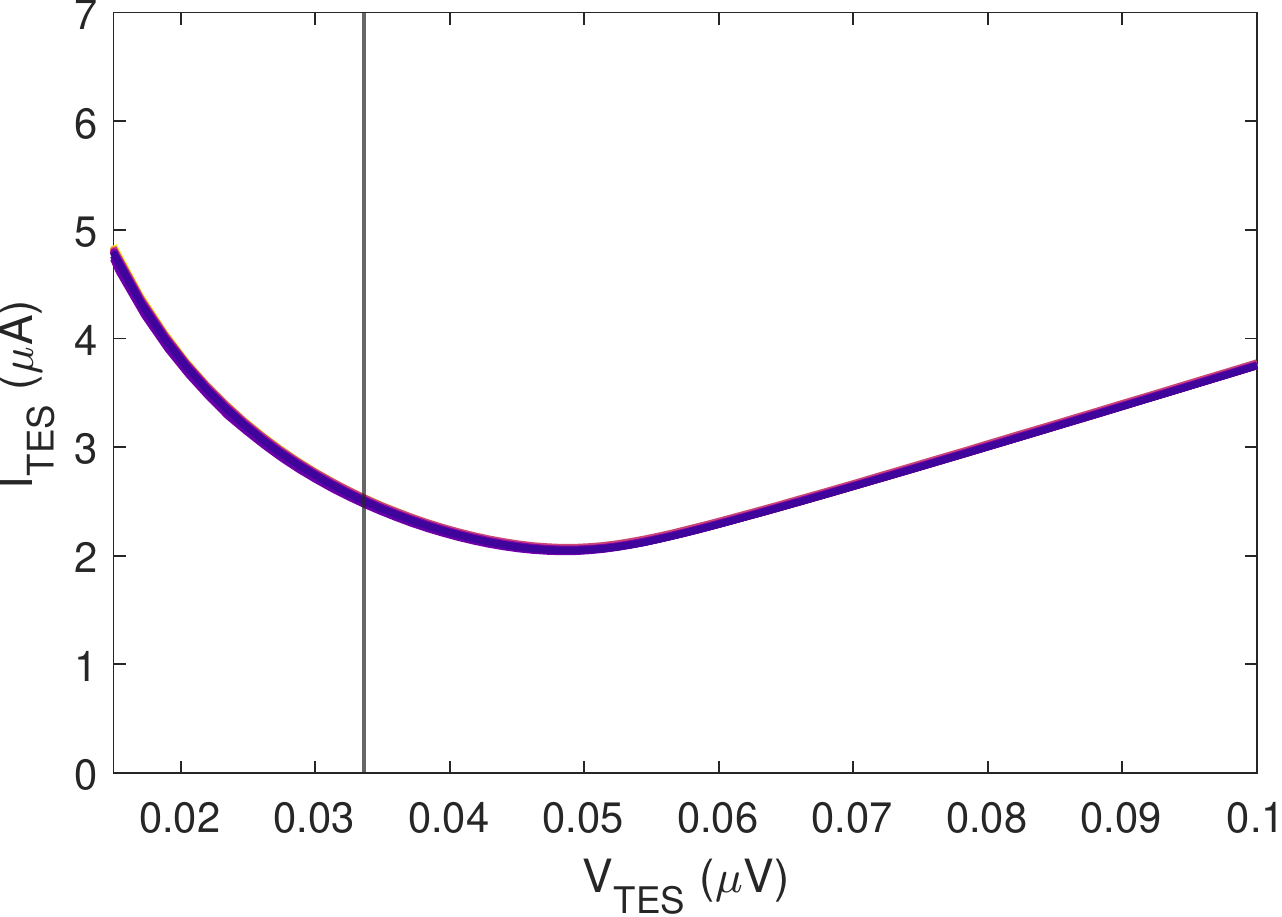}{0pt} && 
    \subfigimg[height=5.5cm]{b)}{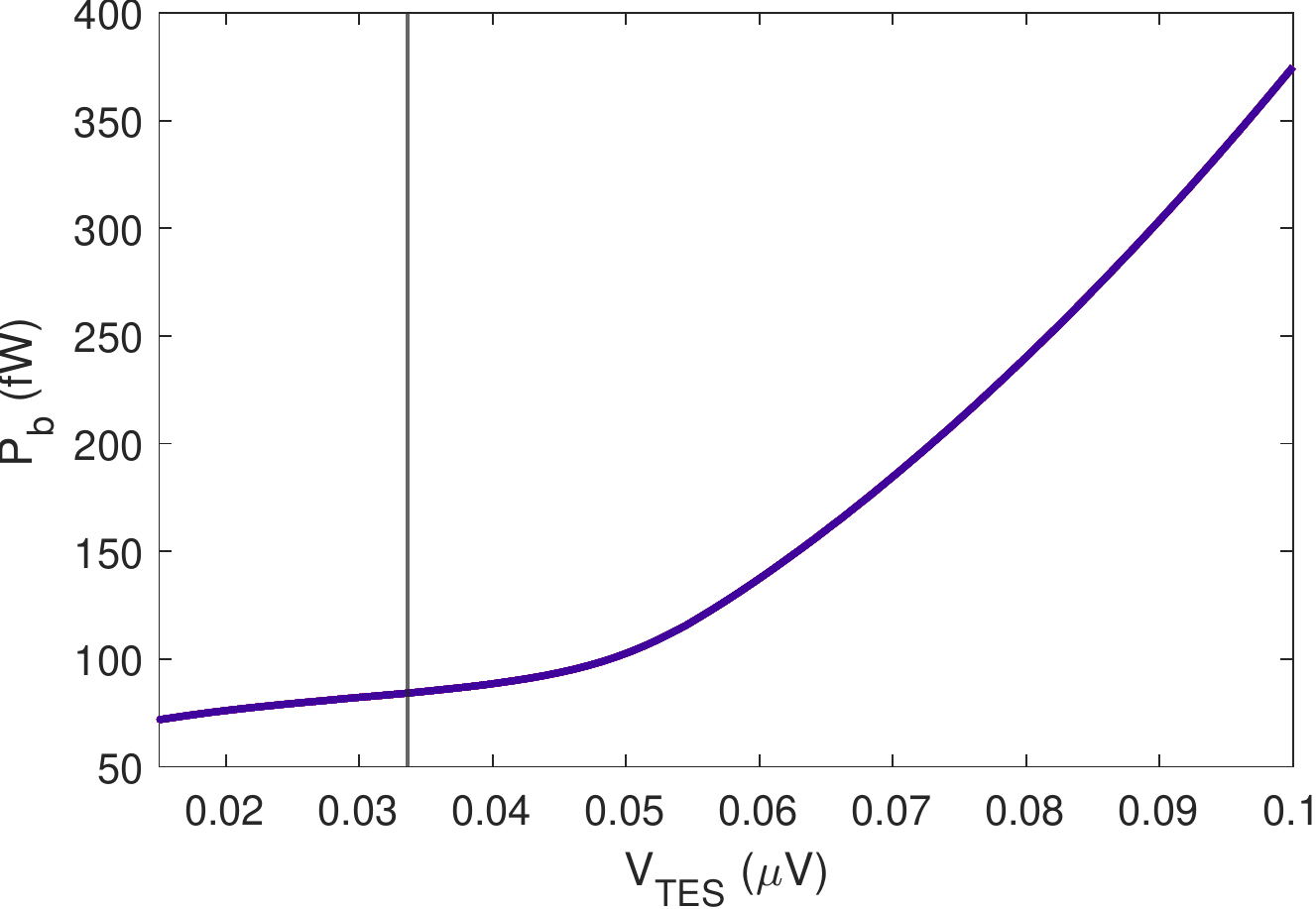}{0pt}
\end{tabular}
\caption{a) TES voltage-current relation measured at 120\,mK bath temperature. b) TES power flow to bath plotted against TES bias voltage measured. In both figures, measurements taken from 0\,V to 20\,V applied plate bias at 1\,V intervals have been overlaid with no clear effect of the plate bias at this scale.}
\label{fig:IV}
\end{figure}
Fig.\,\ref{fig:IV}a shows the TES voltage-current relation measured at 120\,mK with applied bias plate voltages spanning 0 to 20\,V. The TES was biased within its superconducting transition and so the device displayed finite resistance and Joule heating. At equilibrium temperature, with no external energy absorption, Joule heating, \(P_v\), and power dissipation, \(P_b\), from the TES to the bath must balance such that
\begin{equation} \label{eq2}
P_v\left(V_\text{TES},T\right) = \frac{V_\text{TES}^2}{R(T)} = P_b\left(T,T_b\right),
\end{equation}
where \(V_\text{TES}\) is the TES bias voltage, \(R(T)\) is the temperature-dependent TES resistance and \(T_b\) is the bath temperature. Fig.\,\ref{fig:IV}b shows the TES voltage-power relation  calculated using Eq.\,\ref{eq2}. In both of these graphs, the measurements at different electric field strengths from 0 to 90\,\si{\kilo\volt\per\meter} have been overlaid with no clear differences between the lines observable at this scale. 

\begin{figure}[ht]
\centering
  \begin{tabular}[c]{@{}p{0.45\linewidth} @{}p{0.05\linewidth} @{}p{0.45\linewidth}}
    \subfigimg[height=5.8cm]{a)}{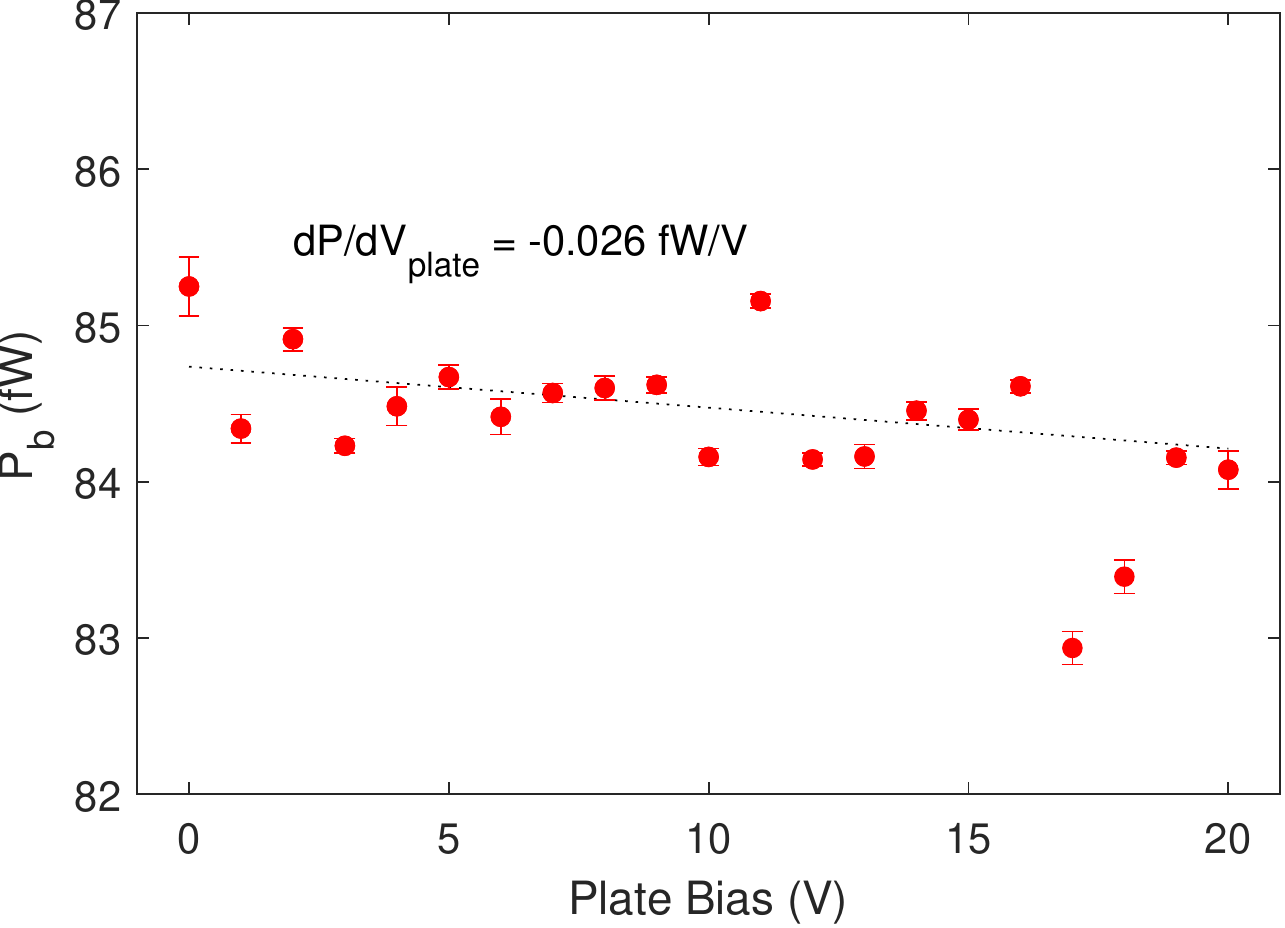}{0pt} &&
    \subfigimg[height=5.8cm]{b)}{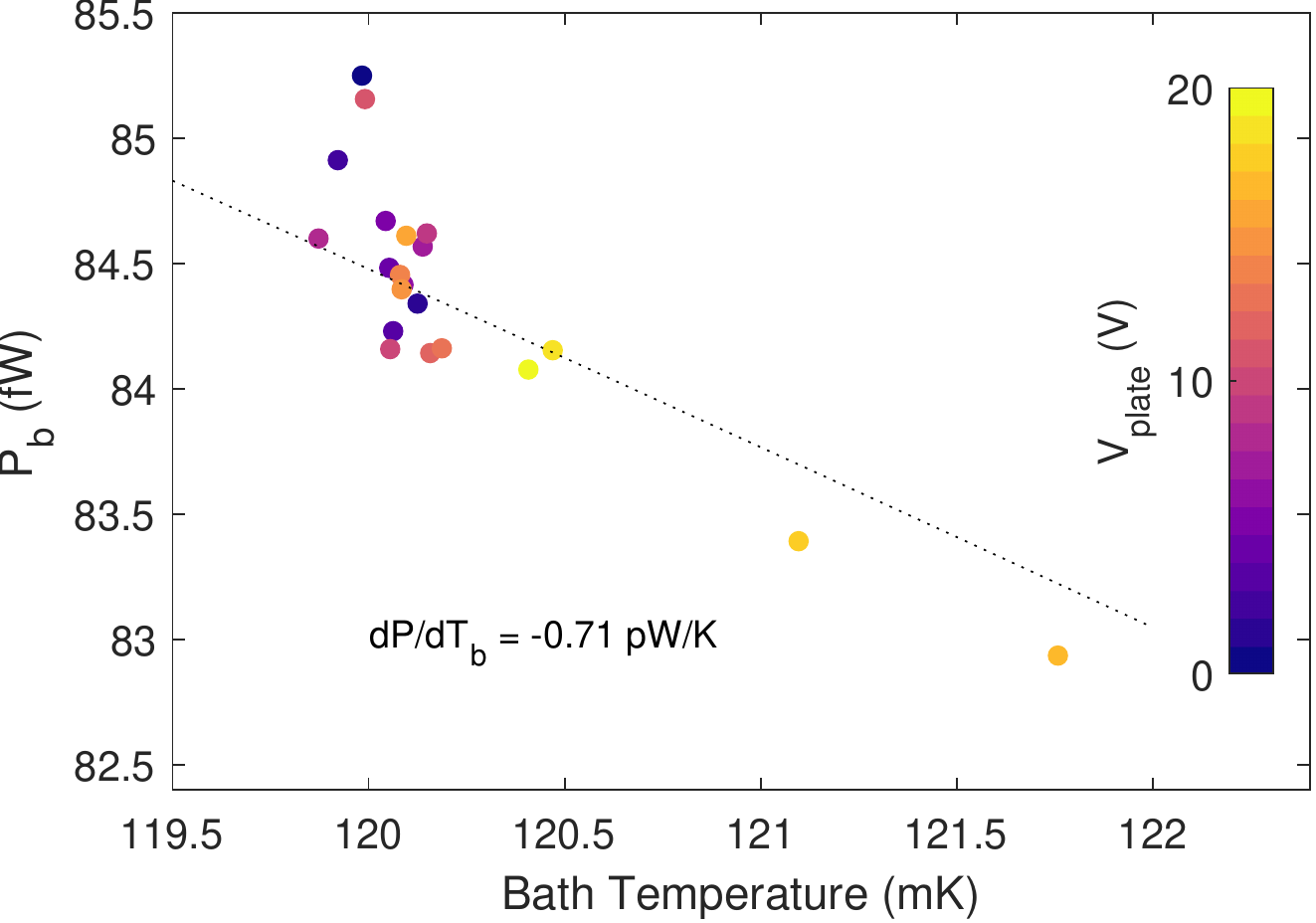}{0pt}
\end{tabular}
\caption{a) TES power flow to bath plotted against plate bias voltage from 0 to 20\,\si{\volt}, measured at the normal-state resistance (\(R_n/2\)) TES bias voltage. The error bars were calculated using a standard deviation of the measured temperature variation during each measurement. b) The data in (a) plotted against bath temperature with the \(V_{plate}\) colour scale showing the applied plate bias.}
\label{fig:Pv}
\end{figure}

Fig.\,\ref{fig:Pv} shows the thermal power dissipation measured at a fixed TES bias point, shown by the vertical lines in Fig.\,\ref{fig:IV} as a function of plate bias voltage (\(V_\text{plate}\)). At this bias point, the TES displays half of its normal state resistance. A small negative correlation was observed between the \(P_b\) and the applied plate bias. Additionally, the measurements at 17 and 18\,V biases were notably smaller than the rest of the measurements. Fig.\,\ref{fig:Pv}b plots the data from Fig.\,\ref{fig:Pv}a but against measured bath temperature where a clear downward trend is also visible. The anomalous readings at 17 and 18\,V are the two datapoints visible over 121\,mK.

The thermal power flow between the TES and the thermal bath can be modelled as
\begin{equation}\label{eq3}
P_b\left(T_c,T_b\right) = K\left(T_c^n - {T_b}^n\right),
\end{equation}
where \(T_c\) is the critical temperature, \(K\) is a constant determining the strength of the thermal conductance between the TES and the bath, and \(n\) is a parameter dependent on the dimensionality of this thermal link. The gradient of the dotted line in Fig.\,\ref{fig:IV}c was calculated using Eq.\,\ref{eq2} and previously measured values of \(n\), and \(K\) of 2.4 and 5.4\,\si{\pico\watt\per\kelvin^n}, respectively. The agreement between predicted and observed temperature dependence shows that the anomalously low measurements at 17 and 18\,V biases are demonstrably dominated by temperature effects and so they were excluded from the gradient calculation in Fig.\,\ref{fig:IV}b, resulting in a gradient of -0.026\(\pm\)0.012\,\si{\femto\watt\per\volt}. The standard error of this linear fit is \(\pm\)0.012\,\si{\femto\watt\per\volt} showing that the TES is far more sensitive to sub-millikelvin changes in bath temperature than the electric field strengths below 90\,\si{\kilo\volt\per\meter} and the observed trend in \(dP/dV\) could have arisen from small drifts in bath temperature below the temperature measurement's precision limit.

The results from Fig.\,\ref{fig:IV} and Fig.\,\ref{fig:Pv} do not show any observable electrostatic effect but can be used to obtain an upper bound on the relation between TES thermal conductivity \(G\) and electric field strength. Thermal conductivity, \(G\) depends on \(K\) by the relation 
\begin{equation}\label{eq3b}
G = \frac{dP}{dT_c} = nKT_c^{n-1}.
\end{equation}
If we assume that \(dP/dV_\text{plate}\) from Fig.\,\ref{fig:Pv}a is solely due to a linear dependence of K due to electric field strength, the upper bound sensitivity of \(G\) to the applied field is calculated as \((dG/dE) = 1\times10^{-19}\)\,(\si{\watt\per\kelvin})/(\si{\volt\per\meter}). This value can be rewritten as a proportional change in \(G\) of \(G^{-1}(dG/dE)\) = -0.07\,\(\left(\si{\mega\volt/\meter}\right)^{-1}\).

\begin{figure}[ht]
\centering
\begin{tabular}[c]{@{}p{0.45\linewidth} @{}p{0.05\linewidth} @{}p{0.45\linewidth}}
    \subfigimg[height=5.8cm]{a)}{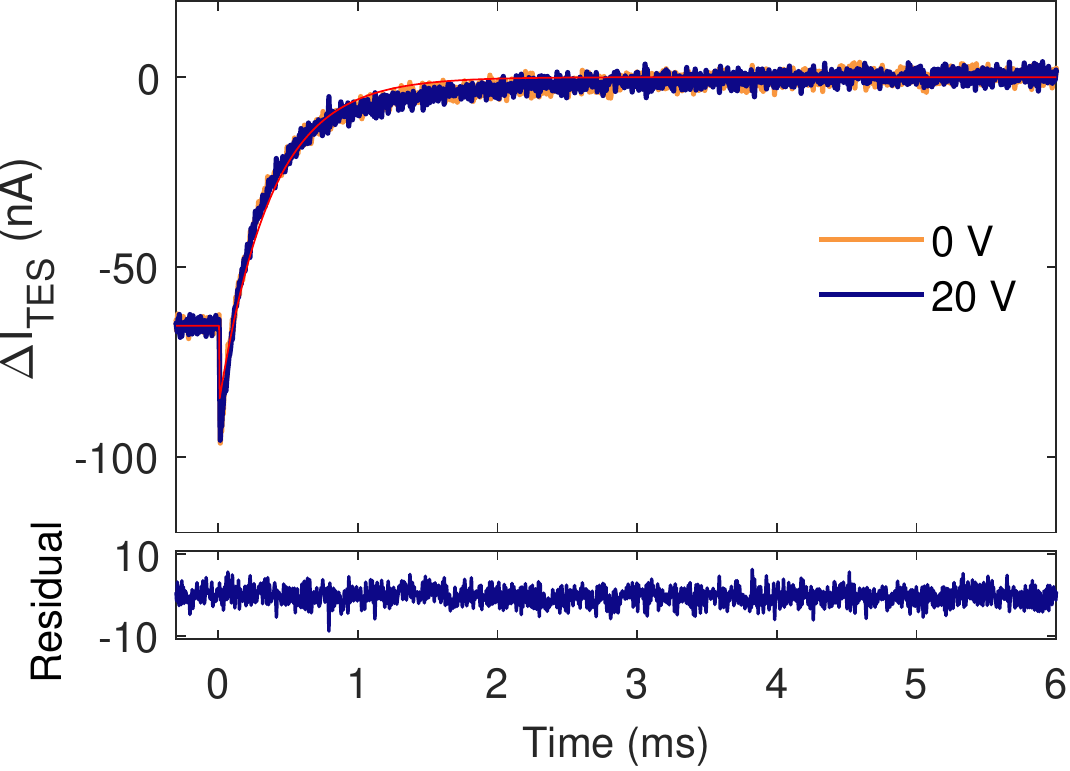}{0pt} &&
    \subfigimg[height=5.8cm]{b)}{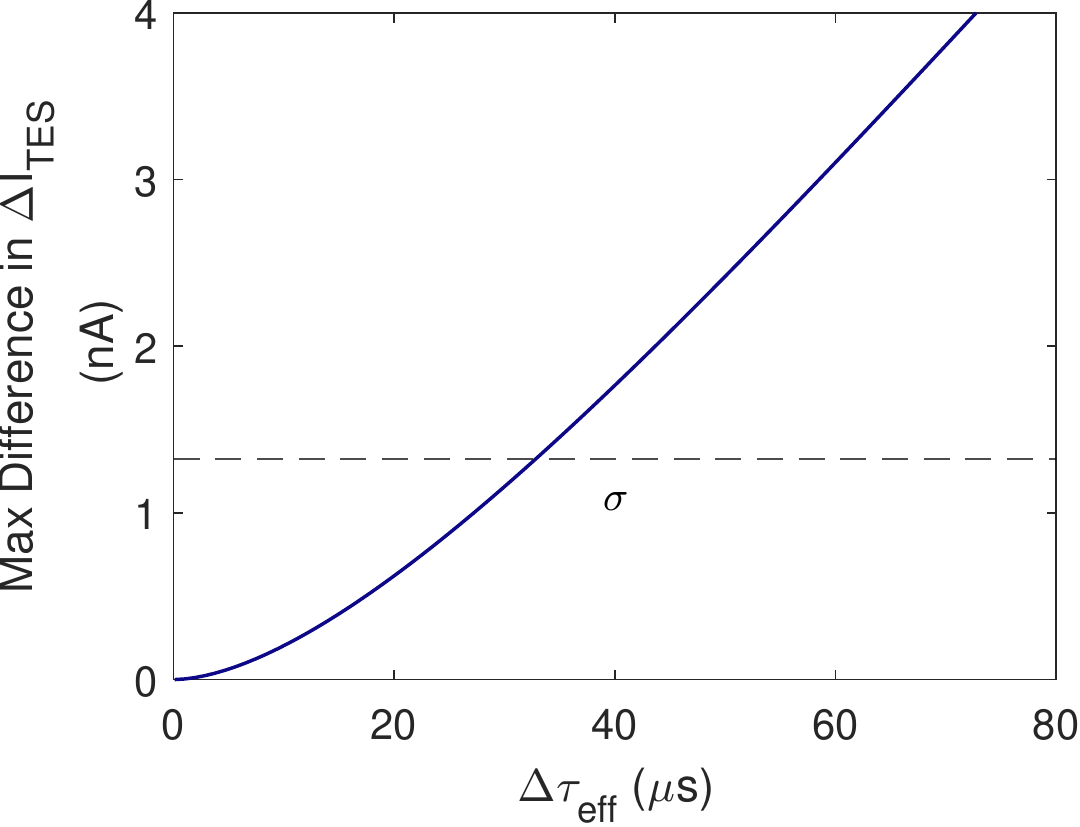}{-5pt}
\end{tabular}
\caption{a) TES response measured at 0 and 20\,\si{\volt} applied plate bias. The inset provides the residual from subtracting the 0\,V and 20\,V measurements. The red line shows the fitted response using a single exponential decay constant. b) The maximum difference between two exponential decays with time constants differing by \(\Delta\tau_\text{eff}\) calculated using Eq.\,\ref{appx1} for the exponential fit in (a).}
\label{fig:Resp}
\end{figure}
With an upper bound placed upon the change in thermal conductivity, the sensitivity of the effective TES response time, \(\tau_{\text{eff}}\), to electric fields places a limit on the change in TES heat capacity \(C\) to electric field strength. The TES effective response time is given by
\begin{equation} \label{eq4}
    \tau_{\text{eff}} = \frac{nC}{\alpha G},
\end{equation}
where \(\alpha\) is a measure of the TES temperature-resistance gradient. Fig.\,\ref{fig:Resp}a shows the measured TES response at 0 and 20\,V plate bias. Both responses were modelled as exponential decays with fitted decay constants of  0.370\,ms and 0.368\,ms. The standard error of these fits are \(\pm\)0.002\,ms, highlighting the similarity in the responses. It is important to note that a single exponential decay constant does not fully characterise the TES response and so additional uncertainty is present in the estimates of \(\tau_{\text{eff}}\) that has not been accounted for. However, of interest here is the possible change in \(\tau_{\text{eff}}\) upon applying an electric field. If the observed time constant \(\tau_{\text{eff}}\) is assumed to differ at 0 and 20\,V applied biases, the deviation in current between the two time response measurements, represented by \(\overline{\Delta I}\), is
\begin{equation} \label{appx1}
    \overline{\Delta I} = A\left(
    \text{exp}\left({ \frac{-t}{ \tau_{\text{eff}} } }\right)
    -\text{exp}\left({ \frac{-t}{ \tau_{\text{eff}}  +\Delta\tau} }\right)
    \right)
\end{equation}
with the difference in response times represented as \(\Delta\tau\), for a pulse starting at \(t=0\) with maximum amplitude \(A\). As no shift in TES response was observed, the magnitude of \(\Delta\tau\) must be constrained by the measurement noise. One conservative estimate of this constraint would be to require that \(\overline{\Delta I}\) be smaller than the observed RMS noise in the measurement. \(\overline{\Delta I}\) reaches its maximum value at
\begin{equation} \label{appx2}
    t_m = \frac{-\tau_{\text{eff}}\left(\tau_{\text{eff}}+\Delta\tau\right)}
    {\Delta\tau}
    \text{log}\left(\frac{\tau_{\text{eff}}}{\tau_{\text{eff}} + \Delta\tau}\right).
\end{equation}
Fig.\,\ref{fig:Resp}b shows the maximum deviation calculated using Eq.\,\ref{appx1} and  Eq.\,\ref{appx2} between two exponential decays of the form fitted in \ref{fig:Resp} when modifying one decay constant. The dashed horizontal line labelled \(\sigma\) represents the RMS noise level in the TES response measurements. The upper bound deviation in \(\tau_{\text{eff}}\) can be seen to be 33\,\si{\micro\second}, corresponding to an 9\% uncertainty in \(\tau_{\text{eff}}\). The constraint on \(\tau_{\text{eff}}\) is an order of magnitude less restricting than that placed on the thermal conductivity using Fig.2 and so can be largely attributed to a possible change in heat capacity. Once more, assuming \(C\) varies linearly with applied field strength, a 9\% change in heat capacity by applying 20\,V plate bias sets an upper bound of \(C^{-1}(dC/dE) = \pm1.0\,\left(\si{\mega\volt/\meter}\right)^{-1}\) on the sensitivity of TES heat capacity to applied electric fields.

\section{\label{sec:level3}Conclusion}
We have measured no observable effect of electric field upon TES behaviour up to  90\,\si{\kilo\volt\per\meter}. This null result has been used to determine upper bounds upon the change of TES heat capacity and thermal conductivity in the presence of a DC electric field. The relative changes in thermal conductivity and heat capacity are smaller than -0.07\,\(\left(\si{\mega\volt/\meter}\right)^{-1}\) and 1.0\,\(\left(\si{\mega\volt/\meter}\right)^{-1}\), respectively. These values are limited by measurement uncertainty rather than any observable effect and so it is likely that significantly greater field strengths can be used with no negligible impact on TES performance. This lack of sensitivity allows for the use of components with a strong DC field to be used in the vicinity of operational TESs without the need for shielding. Such components would be desirable when using TESs in the presence of charged particles, either to screen unwanted particles or enhance absorption efficiency of charge particles of interest.

\acknowledgments 
The authors would like to thank Alexander Shard for very helpful discussions about this work and the motivations behind it. This work was supported by EPSRC Cambridge NanoDTC (EP/L015978/1) and National Measurement System of the UK Department for Science, Innovation and Technology (Project 127126). 
\bibliography{report} 
\bibliographystyle{spiebib} 

\end{document}